\documentclass[prd,a4paper,showpacs,preprint,byrevtex]
{revtex4}
\usepackage{graphicx}
\usepackage{dcolumn}
\usepackage{amsmath}
\usepackage{bm}

\begin{document}

\title{Observer with a constant proper acceleration}

\author{Claude \surname{Semay}}
\thanks{FNRS Research Associate}
\email[E-mail: ]{claude.semay@umh.ac.be}
\affiliation{Groupe de Physique Nucl\'{e}aire Th\'{e}orique,
Universit\'{e} de Mons-Hainaut,
Acad\'{e}mie universitaire Wallonie-Bruxelles,
Place du Parc 20, BE-7000 Mons, Belgium}

\date{\today}

\begin{abstract}
Relying on the equivalence principle, a first approach of the general
theory of relativity is presented using the spacetime metric of an
observer with a constant proper acceleration. Within this non inertial
frame, the equation of motion of a freely moving object is studied
and the equation of motion of a second accelerated observer with the
same proper acceleration is examined. A comparison of the metric of the
accelerated observer with the metric due to a gravitational field
is also performed.
\end{abstract}


\maketitle

\section{Introduction}
\label{sec:intro}

The study of a motion with a constant proper acceleration is a classical
exercise of special relativity that can be found in many textbooks
\cite {sear68,misn73,sema05}. With its analytical solution, it is
possible to show that the limit speed of light is asymptotically reached
despite the constant proper acceleration. The very prominent notion of
event horizon can be introduced in a simple context and the problem of
the twin paradox can also be analysed. In many articles of
popularisation, it is sometimes stated that the point of view of an
observer with a constant proper acceleration cannot be treated within
the theory of special relativity and that theory of general relativity
is absolutely necessary. Actually, this is not true. The point of view
of an uniformly accelerated observer has been studied, for instance, in
Refs.~\cite{sear68,misn73,desl87,boug89}. In this paper, some particular
topics are summed up and developed.

The metric for an observer with a proper constant acceleration (coming
from infinity and coming back towards infinity) is built in
Sec.~\ref{sec:metric} by considering a series of inertial frames
instantaneously at rest with this observer. The notions of local time
and local velocity for such an observer are introduced in
Secs.~\ref{sec:ltime} and \ref{sec:lspeed}. In particular, the equation
of motion of a freely moving object is studied in Sec.~\ref{sec:emffobj}
and its local velocity calculated in Sec.~\ref{sec:vffobj}. The problem
of the distance existing between two accelerated observers with the same
constant proper acceleration \cite{evet72,bell87,tart03} is examined in
Sec.~\ref{sec:distobs}, in terms of the metric associated with one of
the accelerated observer. At last, the metric of the accelerated
observer is compared with the metric due to a gravitational field
in Sec.~\ref{sec:mgravpot}.

The equivalence principle of the general theory of relativity states
that a gravitational field and an acceleration field are locally
equivalent. This means that, in a sufficiently small region of space and
for a sufficiently small duration, the gravitational field can be
cancelled in a suitable accelerated frame. Conversely, a gravitational
field can be simulated by an accelerated frame. Thus, this paper can be
considered as a first approach of the general theory of relativity by
studying the point of view of an accelerated observer.

\section{Metric of an observer with a constant proper acceleration}
\label{sec:metric}

The equation of motion, along a straight line, of an accelerated
observer with a constant proper acceleration can be found in a great
number of textbooks
\cite{sear68,misn73,sema05} (in the following, the coordinates
perpendicular to the velocity of the observer do not play any role and
they are ignored). If $A$ is the modulus of the
proper acceleration, $v' = dx'/dt'$ the velocity of the observer and
$\varphi' = dv'/dt'$ its acceleration along the $x$-axis, both measured
in an inertial
frame $\mathcal{R'}$, the one-dimensional equation of motion is given by
\begin{equation}
\label{2.125}
\varphi' = \left(1-\frac{{v'}^2}{c^2}\right)^{3/2} A.
\end{equation}
With the initial conditions $v'=0$ and $x'=0$ at $t'=0$, the integration
of Eq.~(\ref{2.125}) gives
\begin{eqnarray}
\label{2.126}
v' &=&\frac{A\,t'}{{\left(1+{\displaystyle
\frac{A^2{t'}^2}{c^2}}\right)^{1/2}}}, \\ \label{2.126b}
x' &=& \frac{c^2}{A} \left[\left(1+\frac{A^2{t'}^2}{c^2}\right)^{1/2} -
1\right].
\end{eqnarray}
This motion is also called hyperbolic motion because the last equation
can be recast into the form
\begin{equation}
\label{2.127}
\left(\frac{A\,x'}{c^2} + 1\right)^2 - \left(\frac{A\,ct'}{c^2}\right)^2
=1,
\end{equation}
which is the equation of a branch of hyperbola in spacetime (see
Fig.~\ref{fig:oa1}). The
asymptotes of this curve are the two straight lines with equations
$ct'=\pm (x' +c^2/A)$. The velocity of each physical object is such that
$|dx'/dt'| \le c$. This implies that the angle between the tangent at
each point of the world line of an object and the time-axis in
Fig.~\ref{fig:oa1} is always comprised between 0 and $\pi/4$.
Consequently, the asymptotes of the world line of the accelerated
observer defines two event horizons: The events
``above" the future horizon cannot send any information to the
accelerated observer, who cannot send any information to events located
``below" the past horizon. These two event horizons will be discussed
below.
\begin{center}
\begin{figure}[htb]
\caption{World line of an observer with a constant
proper acceleration $A$, with $x'=0$ and $v'=0$ at $t'=0$ in an inertial
frame $\mathcal{R'}$. \label{fig:oa1}}
\end{figure}
\end{center}

The well known relation between infinitesimal intervals of proper time
and of coordinate time,
\begin{equation}
\label{2.98}
d\tau = dt' \sqrt{1-\dfrac{{v'}^2}{c^2}},
\end{equation}
can be integrated to obtain the relation between the elapsed proper
time $\tau$ for the accelerated observer and the elapsed
time $t'$ for a stationary observer in the inertial frame
\begin{equation}
\label{2.129}
t' = \frac{c}{A} \sinh \left( \dfrac{A\,\tau}{c}\right).
\end{equation}
Clocks for the observers are synchronised such that $\tau=0$ when
$t'=0$.
With relations~(\ref{2.126}), (\ref{2.126b}) and (\ref{2.129}), velocity
and position for the accelerated observer can be computed as a function
of the proper time
\begin{eqnarray}
\label{2.130}
v' &=& c\tanh \left( \dfrac{A\,\tau}{c}\right), \\ \label{2.130b}
x' &=& \frac{c^2}{A}
\left[\cosh\left(\dfrac{A\,\tau}{c}\right)-1\right].
\end{eqnarray}

A system of local coordinates for the accelerated observer can be built
by considering a series of inertial frames instantaneously at
rest with this observer. A particular event on the world line of the
accelerated observer is noted $M$. This observer occupies a position
$x'_M$ at a time $t'_M$ in the inertial frame $\mathcal{R'}$ and is
characterized by a reduced velocity $\beta'_M=v'_M/c
=(A\,t'_M/c)/\sqrt{1+A^2{t'_M}^2/c^2}$ in this frame. A
new inertial frame $\mathcal{R''}$ can be built with its spacetime
origin on $M$ and with a reduced velocity $\beta'_M$ in $\mathcal{R'}$
(see Fig.~\ref{fig:oa2}). The Lorentz transformation between the two
inertial frames $\mathcal{R'}$ and $\mathcal{R''}$ can be written
\begin{eqnarray}
\label{metobs1}
ct'-ct'_M &=& \gamma(v'_M) \left(ct'' + \beta'_M\,x''\right), \nonumber
\\
x'-x'_M &=& \gamma(v'_M) \left(x'' + \beta'_M\,ct''\right).
\end{eqnarray}
Equation~(\ref{2.126}) implies that
\begin{equation}
\label{betam}
\gamma(v'_M)=\sqrt{1+A^2{t'_M}^2/c^2} \quad {\rm and} \quad
\beta'_M\,\gamma(v'_M)=A\,t'_M/c.
\end{equation}
At the event $M$, the spacetime coordinates of the accelerated observer
in the inertial
frame $\mathcal{R''}$ are obviously $x''_M=0$ and $t''_M=0$. By
expressing $x'_M$ as a function of $t'_M$, the system~(\ref{metobs1})
becomes
\begin{eqnarray}
\label{metobs2}
ct'&=& \left( 1 + \frac{A^2\,{t'_M}^2}{c^2} \right)^{1/2} ct'' + ct'_M
\left( 1 + \frac{A\,x''}{c^2} \right), \nonumber \\
1 + \frac{A\,x'}{c^2} &=& \left( 1 + \frac{A^2\,{t'_M}^2}{c^2}
\right)^{1/2} \left( 1 + \frac{A\,x''}{c^2} \right) + \frac{A^2}{c^4}
ct'_M\,ct''.
\end{eqnarray}
\begin{center}
\begin{figure}[htb]
\caption{Building of the instantaneous proper frame $\mathcal{R''}$ of
the accelerated observer. The event $M$ indicates the position of the
observer in spacetime. The temporal coordinate of the event $E$ is
$\tau$ for this observer and is $t''=0$ in the inertial frame
$\mathcal{R''}$. \label{fig:oa2}}
\end{figure}
\end{center}

The accelerated observer can now use the inertial frame $\mathcal{R''}$
to build a proper system of coordinates. If an event $E$ occurs at a
time $t''=0$ in the frame $\mathcal{R''}$, it is natural for the
accelerated
observer to consider that this event occurs at the time $t$ indicated by
a clock of the observer, since the time coordinate of the observer is
also $t''=0$ in the frame $\mathcal{R''}$ (see Fig.~\ref{fig:oa2}). This
time $t$ is then identical to its proper time $\tau$. It is also natural
to assign at this event $E$ a position $x$ in the proper frame of the
observer which is identical to the position $x''$ of $E$ in the frame
$\mathcal{R''}$. Finally, the event $E$ with spacetime coordinates
$(t''=0,x'')$ in the frame $\mathcal{R''}$ has spacetime coordinates
$(t=\tau,x=x'')$ in the proper frame of the accelerated observer. The
relation~(\ref{2.129}) indicates that the proper time of the observer at
time $t'_M$ is given by $A\,t'_M/c=\sinh (A\,\tau/c)$. Consequently, the
factor $\sqrt{1+A^2{t'_M}^2/c^2}$ is equal to $\cosh (A\,\tau/c)$ and
the system~(\ref{metobs2}) can be rewritten
\begin{eqnarray}
\label{apacc6}
ct' &=& \left( x +\dfrac{c^2}{A} \right) \sinh\left(
\dfrac{A\,t}{c}
\right), \nonumber \\
x' &=& \left( x +\dfrac{c^2}{A} \right) \cosh\left(
\dfrac{A\,t}{c}
\right) - \dfrac{c^2}{A}.
\end{eqnarray}
The metric associated with the accelerated observer can now be
determined by computing the invariant $ds^2 = c^2 d{t'}^2 - d{x'}^2$,
which gives
\begin{equation}
\label{etc5}
ds^2 = g(x) c^2 dt^2 - dx^2 \quad {\rm with} \quad
g(x) = \left( 1 + \dfrac{A\, x}{c^2}\right)^2.
\end{equation}
It is worth mentioning that a spacetime with such a metric has
no curvature, since this metric is obtained from a flat metric with a
change of coordinates. It can be verified that the calculation of the
curvature tensor for the metric~(\ref{etc5}) gives a null
scalar curvature, as expected \cite{misn73}.

\begin{center}
\begin{figure}[tbh]
\caption{Coordinate lines for constant time $t$ (straight lines) and for
constant position $x$ (hyperbolas), for an observer with a constant
proper acceleration $A$, with $x'=0$ and $v'=0$ at $t'=0$ in an inertial
frame $\mathcal{R'}$. The world line of the observer is the coordinate
line $x=0$. \label{fig:oa3}}
\end{figure}
\end{center}

In the proper frame of the accelerated observer, called $\mathcal{R}$
here, each object with a position $x$ positive (negative) is located
above (below) the observer, since the acceleration defines a privileged
vertical direction along the $x$-axis. It is worth noting that the
metric~(\ref{etc5})
is not defined for $x \le -c^2/A$. To understand why this position is so
particular, it is necessary to study the spacetime structure of the
surroundings of the accelerated observer. With relations~(\ref{apacc6}),
it is possible to determine the equations of the coordinates lines of
the frame $\mathcal{R}$ in the inertial frame $\mathcal{R'}$. In this
last frame, the equation of a coordinate line $t$ with $x=x_0$ constant
is
\begin{equation}
\label{lctemps}
\left( x' +\frac{c^2}{A} \right)^2 - c^2 {t'}^2 =
\left( x_0 +\frac{c^2}{A} \right)^2.
\end{equation}
This curve is a branch of hyperbola whose asymptotes are the two event
horizons mentioned above. These horizons are located on the degenerate
asymptotes obtained with $x_0=-c^2/A$ in Eq.~(\ref{lctemps}). Obviously,
the world line of the accelerated observer in the frame $\mathcal{R'}$
is given by $x_0=0$. The equation of a coordinate line $x$ with $t=t_0$
constant is
\begin{equation}
\label{lcxpos}
ct' = \tanh\left( \frac{A\,t_0}{c}\right) \left( x' +\frac
{c^2}{A} \right).
\end{equation}
This is a straight line containing the event $(x',ct')=(-c^2/A,0)$, that
is to say the intersection of the two event horizons. When
$t_0\rightarrow \infty$, the straight line is parallel to the
bisector of the axis $x'$ and $ct'$. Some coordinate lines are drawn in
Fig.~\ref{fig:oa3}. It can be seen that the future horizon and the past
horizon correspond respectively to the time coordinate lines
$t=+\infty$ and $t=-\infty$. Both horizons form also the space
coordinate line $x=-c^2/A$. The spacetime region with $x \le -c^2/A$ is
then behind the event horizons.

\begin{center}
\begin{figure}[htb]
\caption{Photons, emitted during a finite time $\Delta t'$ by a
stationary observer ``below" the future
horizon in the inertial frame $\mathcal{R'}$, can reach the uniformly
accelerated observer after an infinite time. \label{fig:oa4}}
\end{figure}
\end{center}

The event horizons are then always located at a distance $c^2/A$
downward for the accelerated observer. In order to understand the
nature of these horizons, let us assume that a stationary observer in
the inertial frame $\mathcal{R'}$ sends photons to the accelerated
observer: The emission starts in order that the first photons reach the
accelerated observer at $t=t'=0$, for instance; It stops when the
stationary observer crosses the future horizon (photons emitted later
cannot be received). It is clear from Fig.~\ref{fig:oa4} that all
photons are received after an infinite time, despite the fact that they
are emitted during a finite time $\Delta t'$.

\section{Local time}
\label{sec:ltime}

Let us consider a clock at rest at position $x$ in the
proper frame $\mathcal{R}$ of the accelerated observer. An interval of
proper time $d\tau$ for this clock corresponding to an interval of
coordinate time $dt$ is given by
\begin{equation}
\label{etc5bis}
d\tau=\sqrt{g(x)}\,dt = \left( 1 + \dfrac{A\, x}{c^2}\right) dt,
\end{equation}
since $dx=0$. Let us remark that
the coordinate time is also the proper time at the level of the
accelerated observer ($x=0$). The time flows with the same rate for all
clocks with
the same ``altitude" in $\mathcal{R}$ (same value of $x$). On the
contrary,
clocks located at different altitudes measure different intervals of
proper time for a same interval of coordinate time. The time flows more
slowly for all clocks located below a reference clock. In particular,
the interval
of proper time $d\tau$ vanishes at the level of event
horizons. For the accelerated observer, the time ``freezes" at the
coordinate $x=-c^2/A$ (see Fig.~\ref{fig:oa4}).

It is worth noting that, for a value of $A$ close to the terrestrial
gravitational acceleration (around 10~m/s$^2$), $c^2/A$ is around one
light year (10$^{16}$~m). With this value of $A$, two clocks located 1~m
apart will be get out of synchronisation by about 1~s every 10$^{16}$~s
(around $3\times 10^8$~yr).

An interval of local time can be defined, $d\tilde t = \sqrt{g(x)}\,dt$,
with which it is possible to build in the vicinity of a particular event
a local metric similar to the usual Minkowski metric
\begin{equation}
\label{etc6}
ds^2 = c^2 d\tilde t\,^2 - dx^2.
\end{equation}
This metric can be considered as the metric of an inertial frame
instantaneously at rest with the proper frame of the accelerated
observer. But, in this frame, the
intervals of time considered must be small enough in order that the
accelerated frame do not move appreciably. An
interval of local time corresponds then to an interval
of time measured by a clock at rest in this instantaneous inertial
frame. The metric~(\ref{etc6}) allows only a local
description of an infinitesimal region of spacetime like a Minkowski
spacetime with local coordinates $(x,\tilde t)$.

\section{Local velocity}
\label{sec:lspeed}

Let us consider two events $E_1$ and $E_2$ on the world line of an
object
that moves in the non inertial frame $\mathcal{R}$. The coordinate time
interval between these events is $dt$ and the coordinate distance is
$dx$. It is possible to define three types of velocity for this object:
\begin{itemize}
\item The coordinate velocity $v=dx/dt$.
\item The proper velocity $u = dx/d\tau$, which is the spatial part of
the world velocity.
\item The local velocity $\tilde v=dx/d\tilde t$, computed with the
interval of local time $d\tilde t=\sqrt{g(x)}\,dt$.
\end{itemize}
Hence, we have
\begin{equation}
\label{etc8}
v = \sqrt{g(x)}\, \tilde v,
\end{equation}
and
\begin{equation}
\label{etc9}
u = \dfrac{\tilde v}{\sqrt{1-{\tilde v}^2/c^2}} =
\dfrac{v}{\sqrt{g(x)-v^2/c^2}}.
\end{equation}
The local velocity $\tilde v$ is the coordinate velocity of the object
in the inertial frame that is instantaneously at rest relative to the
non inertial frame of the accelerated observer (see previous section).
This velocity cannot thus exceed the speed of light. This is not the
case for the coordinate velocity $v$ since, according to
Eq.~(\ref{etc8}), $v$ could exceed $c$ if $g(x)$ is large enough. If the
object is a photon, its local velocity $\tilde v$ must be the invariant
$c$, but its coordinate velocity varies as a function of its position.

\section{Equation of motion of a freely moving object}
\label{sec:emffobj}

The interval of proper time elapsed between two events infinitesimally
close on the world line of a moving object in the non inertial frame is
given by
\begin{equation}
\label{etc10}
d\tau = \dfrac{1}{c} \sqrt{g(x)\,c^2dt^2-dx^2} =
\sqrt{g(x)-v^2/c^2}\, dt,
\end{equation}
where $v$ is the coordinate velocity of the object. The finite duration
of proper time between two events $E_1$ and $E_2$ on this world line is
calculated by integration
\begin{equation}
\label{etc11}
\Delta \tau = \int^{E_2}_{E_1} d\tau = \int^{E_2}_{E_1}
\sqrt{g(x(t))-v^2/c^2}\, dt.
\end{equation}

The principle of equivalence requires that the world line of a freely
moving object that passes through two fixed events is such that the
elapsed proper time between these two events is maximum
\cite{sear68,misn73}. If the integral~(\ref{etc11}) is
written
\begin{equation}
\label{etc12}
\Delta \tau = \int^{E_2}_{E_1} L(t)\, dt\quad {\rm with} \quad L(t) =
\sqrt{g(x(t))-v^2/c^2},
\end{equation}
then the calculus of variations shows that the functional $L(t)$ must
satisfy the following condition
\begin{equation}
\label{etc13}
\dfrac{d}{dt} \dfrac{\partial L}{\partial v} -
\dfrac{\partial L}{\partial x} = 0.
\end{equation}
This is the Euler-Lagrange equation, which ensures that the quantity
$\Delta \tau$ is extremum.

Let us mention that this problem is formally identical to the
computation of the minimal length between two points on a curved space
described by a spatial metric. The curve satisfying this condition is
called a geodesic. Here, the curve in space is replaced by a world line
in the spacetime and the variation produces a maximum in
the lapse of proper time instead of a minimum of spatial distance.

Substituting $L(t)$ by its value in Eq.~(\ref{etc13}) gives
\begin{equation}
\label{etc14}
\dfrac{1}{c^2} \dfrac{d^2 x}{dt^2} = -\dfrac{1}{2}
\dfrac{dg}{dx}\left(1-\dfrac{2\,v^2}{c^2g}\right),
\end{equation}
by using $\dfrac{dg}{dt} =  \dfrac{dg}{dx}\dfrac{dx}{dt}
= v\dfrac{dg}{dx}$. The quantity $d^2x/dt^2$ is the coordinate
acceleration of the moving object in the non inertial frame of the
accelerated observer. Let us remark that $d^2x/dt^2=0$, as expected, in
the case of an inertial frame for which $g(x)=1$.

With the function $g(x)$ given by Eq.~(\ref{etc5}), the coordinate
acceleration is written
\begin{equation}
\label{etc14b}
\dfrac{d^2x}{dt^2} = -A \left( 1 + \dfrac{A\, x}{c^2}\right)
\left[ 1 - 2 \left( \dfrac{v/c}{1+A\,x/c^2} \right)^2 \right].
\end{equation}
When the object moves slowly ($v \ll c$) near the accelerated observer
($x \ll c^2/A$), this equation of motion reduces to
\begin{equation}
\label{etc14c}
\dfrac{d^2x}{dt^2} = -A.
\end{equation}
The object accelerates downward at the rate $A$, as expected. This
corresponds to a free fall in a constant gravitational field
(see Sec.~\ref{sec:mgravpot}).

\section{Velocity of a freely moving object}
\label{sec:vffobj}

Equation~(\ref{etc14}) can be integrated to yield the velocity of the
freely moving object as a function of its position in the non inertial
frame. Multiplying both sides of this equation by $dx$ and using the
relations
\begin{equation}
\label{etc16}
{dv\over{dt}} dx = {dx\over{dt}} dv = v\,dv,
\end{equation}
the acceleration equation becomes
\begin{equation}
\label{etc17}
2\,v\,dv = c^2\left(\dfrac{2\,v^2}{c^2g} - 1\right)dg.
\end{equation}
With the notation $w=v^2/c^2$, this equation can be put into the form
\begin{equation}
\label{etc17bis}
\frac{dw}{dg}=\frac{2w}{g}-1.
\end{equation}
This differential equation can be solved by using the separation of
variables method, if one introduces the new function $y=w/g$. But it is
simpler to try a solution of the form
\begin{equation}
\label{etc17ter}
w=\sum_{n=0}^{\infty} a_n\, g^n.
\end{equation}
It is then easy to see that the coefficients $a_n$ are such that
$a_0=0$, $a_1=1$, $a_2$ is
arbitrary and $a_m=0$ for $m \geq 3$. So, we can write
\begin{equation}
\label{etc18}
v^2=c^2g(1-kg),
\end{equation}
where $k$ is the constant of
integration ($k=-a_2$ in Eq.~(\ref{etc17ter})). Using Eq.~(\ref{etc8}),
the local velocity of the freely moving object is given by
\begin{equation}
\label{etc19}
\tilde v^2=c^2(1-kg).
\end{equation}

With this equation, it is possible to prove that the local velocity of
an object cannot exceed the speed of light if its local velocity is
below $c$ at one point of its world line. Since $g(x)$ is positive,
$\tilde v$ can exceed $c$ only if $k$ is negative. Eq.~(\ref{etc19})
implies that
\begin{equation}
\label{etc20}
k = \dfrac{1-\tilde v^2/c^2}{g(x)}.
\end{equation}
If we assume that $\tilde v < c$ at one point of its world line, then
$k > 0$. Since $k$ is a constant, it is always positive and $\tilde v$
is always less than $c$. Alike, if the local velocity of a photon is
equal to $c$ at one point of its world line, then $k=0$. Consequently,
$k$ is null everywhere on the world line and the local velocity of the
photon is always equal to $c$. These results are important since they
show that the local velocity of an object cannot exceed the speed of
light and that the local velocity of a photon is always $c$.

The local velocity of a freely moving object has been calculated in
the proper frame of the accelerated observer. It is possible to compute
this speed from the equation of motion of the object given in the
inertial frame $\mathcal{R'}$. The two approaches must obviously give
the same result. With Eqs.~(\ref{apacc6}), the function $g(x)$ is given
by
\begin{equation}
\label{oaplus1}
g(x)=\left( 1 + \frac{A\,x}{c^2} \right)^2 =
\left( 1 + \frac{A\,x'}{c^2} \right)^2 - \frac{A^2 {t'}^2}{c^2}.
\end{equation}
By differentiating both Eqs.~(\ref{apacc6}) and by noting $v'=dx'/dt$
and $v=dx/dt$, the relation between the two velocities in the two
frames is given by
\begin{equation}
\label{oaplus2}
\frac{v_0'}{c}=\frac{\dfrac{v}{c}+ \left( 1 +
\dfrac{A\,x}{c^2}
\right) \tanh \left( \dfrac{A\,t}{c} \right)}{\dfrac{v}{c} \tanh \left
(\dfrac{A\,t}{c} \right) + \left( 1 + \dfrac{A\,x}{c^2} \right) },
\end{equation}
where $v_0'=v'$ is a constant since the object is moving freely.
The quantity $\tanh (A\,t/c)$ can be rewritten as a function of
coordinates $x'$ and $t'$ using Eqs.~(\ref{apacc6}). After
inversion of Eq.~(\ref{oaplus2}) and using the definition~(\ref{etc8})
of the local speed, it is found that
\begin{equation}
\label{oaplus3}
\frac{\tilde v}{c}=\frac{\left( 1 + \dfrac{A\,x'}{c^2}
\right)
\dfrac{v'_0}{c}- \dfrac{A\,t'}{c}}{\left( 1 + \dfrac{A\,x'}{c^2} \right)
- \dfrac{v'_0}{c} \dfrac{A\,t'}{c} }.
\end{equation}
When $t'=0$, we have $\tilde v=v'_0$, as expected since the proper
frame of the accelerated observer coincides instantaneously with the
inertial frame $\mathcal{R'}$ at this time.

Since the object is freely moving, its equation of motion can be written
in the inertial frame $\mathcal{R'}$ as
\begin{equation}
\label{oaplus4}
x'=x'_0 + v'_0 ( t'-t'_0),
\end{equation}
where $x'_0$ and $t'_0$ are constants. The introduction of this
relation in Eq.~(\ref{oaplus3}) gives the local velocity $\tilde v$
as a function of time $t'$ of the inertial frame, but it is more
interesting to modify Eq.~(\ref{oaplus3}) in order to recover
Eq.~(\ref{etc19}). With the notation
\begin{equation}
\label{oaplus5}
\alpha_0 = 1 + \frac{A}{c^2} (x'_0 - v'_0\,t'_0),
\end{equation}
and with the use of Eqs.~(\ref{oaplus1}) and (\ref{oaplus4}), the
function $g$ for the freely moving object can be written
\begin{equation}
\label{oaplus6}
g=\alpha_0^2 + 2\frac{A\,\alpha_0\,v'_0}{c^2} t' -
\left(1-\frac{{v'_0}^2}{c^2} \right) \frac{A^2}{c^2} {t'}^2.
\end{equation}
With some calculations, it can then be shown, with Eqs.~(\ref{oaplus3}),
(\ref{oaplus4}) and (\ref{oaplus6}), that the local velocity of
this object is given by
\begin{equation}
\label{oaplus7}
\frac{\tilde v^2}{c^2}=1-\frac{1-{v'_0}^2/c^2}{\alpha_0^2} g.
\end{equation}
This equation is identical to Eq.~(\ref{etc19}) and the value of $k$ is
then
\begin{equation}
\label{oaplus7bis}
k=\frac{1-{v'_0}^2/c^2}{\alpha_0^2}.
\end{equation}
A simpler procedure exists to find the value of $k$. Since it is a
constant, it can be evaluated at any time, for instance at $t'=0$. This
constant is then given by formula~(\ref{etc20}) in which $\tilde
v=v'_0$, as explained above, and $g=g(t'=0)=\alpha_0^2$. So the
formula~(\ref{oaplus7bis}) is obtained directly.

From these formulas, it is clear that $\tilde v = v'_0 = \pm c$ for a
photon, as expected. On the event horizons ($x=-c^2/A$), the function
$g$ is
vanishing and the velocity of the object tends towards the speed of
light. From the point of view of the accelerated observer, he is at
infinity when the object reaches a horizon (see
Sec.~\ref{sec:metric} and Fig.~\ref{fig:oa3}), that
is to say when its velocity tends towards the speed of light in the
inertial frame $\mathcal{R'}$. The function $g$ in Eq.~(\ref{oaplus6})
is a quadratic form in $t'$ which possesses a maximum equal to
$\alpha_0^2/(1-{v'_0}^2/c^2)$. In this case, the local velocity is
vanishing. The object and the accelerated observer have then the same
velocity in the inertial frame $\mathcal{R'}$.

Finally, the relation~(\ref{oaplus3}) can be rewritten into the form
\begin{equation}
\label{oaplus3b}
\frac{\tilde v}{c}=\frac{\dfrac{v'_0}{c}-\dfrac{\nu}{c}}
{1- \dfrac{v'_0\,\nu}{c^2}} \quad {\rm with}
\quad \nu = \frac{A\,t'}{{\left(1+{
\dfrac{A\, x'}{c^2}}\right)}}.
\end{equation}
Since $\tilde v$ is only defined for $x' > -c^2/A$ and
$|A\,t'|/c < (1+A\,x'/c^2)$ (see
Fig.~\ref{fig:oa3}), then $|\nu| < c$.
Equation~(\ref{oaplus3b}) is thus the relativistic addition of two
velocities below the speed of light. It is shown again that the local
velocity $\tilde v$ cannot exceed $c$.

It is possible to give an interpretation of the speed $\nu$. The
equalities $\sinh(A\,t/c) = \sinh(A\,\tau/c) = A\,t'_M/c$, where
$t'_M$ is the time coordinate of the accelerated observer in the
inertial frame $\mathcal{R'}$ (see Sec.~\ref{sec:metric}), imply
that the first equation of the system~(\ref{apacc6}) can be rewritten
\begin{equation}
\label{oaplus8}
g^{1/2}= \frac{t'}{t'_M}.
\end{equation}
Using the Eqs.~(\ref{oaplus1}) and (\ref{oaplus8}) in the
formula~(\ref{oaplus3b}) for $\nu$, one can find
$\nu=v'_M$.
The local speed $\tilde v$ of the freely moving object in the proper
frame
of the accelerated observer at coordinate time $t$ is then the relative
speed between the constant velocity $v'_0$ of the object in
the inertial frame $\mathcal{R'}$ and the instantaneous velocity $v'_M$
of the accelerated observer in the same inertial frame at time $t'_M$,
with $A\,t'_M/c = \sinh(A\,t/c)$.

\section{Distance between two observers with a constant proper
acceleration}
\label{sec:distobs}

Let us consider two spaceships with the same proper acceleration $A$
moving in an inertial frame $\mathcal{R'}$. Their world lines, plotted
in Fig.~\ref{fig:oa5}, are such that, at $t'=0$, they are a distance $L$
apart, at rest in $\mathcal{R'}$. At each time in the
inertial frame, the ships possess the same velocity and are the same
distance $L$ apart. A rod of length $L$ is fixed between the two
spaceships when they are at rest. As the speed of the ships increases,
the Lorentz-FitzGerald contraction will occur from
the point of view of stationary observers in $\mathcal{R'}$. This rod
will then tend to be lengthened by an increasing stress since the
distance between the two ships is constant in $\mathcal{R'}$. This
classical problem is studied in many papers (see for instance
Refs.~\cite{evet72,bell87,tart03}).

\begin{center}
\begin{figure}[htb]
\caption{World lines (in bold), in an inertial frame, of two observers
with the same proper acceleration
$A$. The distance between the two observers is constant in the
inertial frame ($L$) but increases in the proper frame of the first
observer ($X(t)$).
\label{fig:oa5}}
\end{figure}
\end{center}

It is interesting to analyse this problem from the point of view of an
observer on board a ship. The parametric equations of motion of the
first ship, as a
function of its proper time $\tau$, in the inertial frame $\mathcal{R'}$
are (see Eqs.~(\ref{2.129}) and (\ref{2.130b}))
\begin{equation}
\label{2mob1}
t' = {c\over A} \sinh \left( \dfrac{A\tau}{c}\right) \quad {\rm and}
\quad x' = {c^2\over A}
\left[\cosh\left(\dfrac{A\tau}{c}\right)-1\right].
\end{equation}
The ones for the second ship in the same frame, as a function of its
proper time $\eta$, are
\begin{equation}
\label{2mob2}
t' = {c\over A} \sinh \left( \dfrac{A\eta}{c}\right) \quad {\rm and}
\quad x' - L= {c^2\over A}
\left[\cosh\left(\dfrac{A\eta}{c}\right)-1\right].
\end{equation}
Clocks in all frames are synchronised: $t'= \tau = \eta = 0$.  Let us
note $X$ the position of the second spaceship in the proper frame of
the first one. The relations between the coordinates in the frame
$\mathcal{R'}$ and the coordinates in the proper frame of the first
spaceship are given by Eqs.~(\ref{apacc6}). By introducing the
coordinates~(\ref{2mob2}) of the second ship in Eqs.~(\ref{apacc6}),
the following relations between the proper times $\tau$ and $\eta$ are
found
\begin{eqnarray}
\label{2mob3}
{c^2\over A} \sinh \left( \dfrac{A\eta}{c}\right) &=& \left(
X +\dfrac{c
^2}{A} \right) \sinh\left(
\dfrac{A\,t}{c}
\right), \nonumber \\
{c^2\over A}
\cosh\left(\dfrac{A\eta}{c}\right) + L &=& \left( X +\dfrac{c
^2}{A} \right) \cosh\left(
\dfrac{A\,t}{c}
\right),
\end{eqnarray}
with the identification $t=\tau$ in Eqs.~(\ref{2mob3}). The position $X$
can be obtained by eliminating the proper time $\eta$ in these
equations. This yields a quadratic equation in $X$ whose the only
physical solution is
\begin{equation}
\label{2mob4}
X(t)=L \cosh\left(\dfrac{A\,t}{c}\right) +
\sqrt{L^2 \sinh^2\left(\dfrac{A\,t}{c}\right) + \dfrac{c^4}{A^2}}
- \dfrac{c^2}{A},
\end{equation}
with $X(0)=L$ as expected. When the proper time
$\tau \rightarrow \infty$, Eq.~(\ref{2mob4}) reduces to
\begin{equation}
\label{2mob5}
X(t) \approx L \cosh\left(\dfrac{A\,t}{c}\right) + L \sinh\left
(\dfrac{A
\tau}{c}\right) = L \exp \left(\dfrac{A\,t}{c}\right).
\end{equation}
From the point of view of an observer in the first spaceship, the
distance increases between the two ships. This in agreement with the
reasoning sustained above about the rod.  This phenomenon is due to the
fact that the spaces of simultaneity for an observer on board the first
space ship are different from the spaces of simultaneity for a
stationary observer in the inertial frame. This can be seen on an
example in Fig.~\ref{fig:oa5} where $L$ is fixed at 0.5~$c^2/A$: At the
time $t=0.6\ c/A$, $X(t)$ is clearly greater then $L$.

\section{Metric and gravitational potential}
\label{sec:mgravpot}

In the Newtonian theory of gravity, the gravitational field is derivable
from a function $\phi(x)$ call the gravitational potential. In a
one-dimensional space, the acceleration $d^2x/dt^2$ of a freely falling
object in an arbitrary frame is given by
\begin{equation}
\label{etc21}
\dfrac{d^2x}{dt^2} = -\dfrac{d\phi}{dx}.
\end{equation}

In the vicinity of an accelerated observer, objects seems to undergo the
effects of a gravitational field (see Eq.~(\ref{etc14c})). One can ask
what kind of relation could exist between this pseudo gravitational
field and the metric of the accelerated observer.
Eq.~(\ref{etc21}) has a form similar to Eq.~(\ref{etc14}), which relates
the acceleration with the spacetime metric. At the classical limit, low
velocity ($v \ll c$) and near Minkowski metric ($g \approx 1$),
Eq.~(\ref{etc14}) reduces to
\begin{equation}
\label{etc22}
\dfrac{d^2x}{dt^2} =  -\dfrac{c^2}{2}\dfrac{dg}{dx}.
\end{equation}
With this approximation, we have
\begin{equation}
\label{etc23}
\dfrac{dg}{dx} = \dfrac{2}{c^2} \dfrac{d\phi}{dx}.
\end{equation}
By integration, it is found that
\begin{equation}
\label{etc24}
g(x) = \dfrac{2}{c^2} \phi(x) + g_0,
\end{equation}
where $g_0$ is a constant of integration allowing to fix the value of
the potential at a particular point. Equation~(\ref{etc14c}), obtained
at the same classical limit, implies that the acceleration is constant
and
equal to $-A$. In this case, the potential is $\phi(x)=A\,x$ (a constant
$\phi_0$ can be absorbed into the constant $g_0$), and
Eq.~(\ref{etc14c}) gives
\begin{equation}
\label{etc25}
g(x) = 1 + \dfrac{2\,A\,x}{c^2}.
\end{equation}
The constant is fixed at $g_0=1$, in order that the metric is a
Minkowski metric at the level of the accelerated observer ($x=0$).

In the general case, the function $g(x)$ is actually given by
\begin{equation}
\label{etc26}
g(x) = \left( 1 + \dfrac{A\, x}{c^2}\right)^2 = 1 + \dfrac{2\,A\,
x}{c^2} + \left(\dfrac{A\, x}{c^2}\right)^2.
\end{equation}
The real metric and the metric at the classical limit coincide
when $x \ll c^2/A$, that is to say for small accelerations or for short
distances from the accelerated observer, as expected.

To conclude let us show with a realistic example that a gravitational
field is locally equivalent to an acceleration field. If an observer is
far form a source of gravitation and travels only on short distances, he
can consider the gravitational field as uniform and can define a
constant
$\gamma$ which measures the value of the local acceleration. Let us
consider the Schwarzschild metric for a mass $M$ with spherical
symmetry,
located at the origin. In the usual spherical coordinates, it is written
\begin{equation}
\label{etc3}
c^2d\tau^2 = \left(1-\dfrac{2\,G\,M}{c^2r}\right)c^2dt^2 -
\dfrac{dr^2}{\left( 1-\dfrac{2\,G\,M}{c^2r}\right)}
- r^2(d\theta^2+\sin^2\theta\, d\phi^2),
\end{equation}
where $G$ is the universal gravitational constant.
For an observer at rest ($dr = d\theta = d\phi = 0$) located at
altitude $r$, the interval of proper time $d\tau$ is related to an
interval of coordinate time $dt$ by
\begin{equation}
\label{etc27}
d\tau^2 = \left( 1-\frac{2\,G\,M}{c^2r} \right) dt^2.
\end{equation}
Let us assume that this observer is located at a great distance $R$ from
the origin ($2\,G\,M/(c^2 R) \ll 1$). He can define a local constant
value $\gamma$ of the acceleration by the standard formula
\begin{equation}
\label{etc28}
\gamma=\frac{G\,M}{R^2}.
\end{equation}
Moreover, this observer makes experiments only for values of $r$ close
to $R$. Thus, he considers values of $r=R+x$ with $x \ll R$. Using
Eq.~(\ref{etc27}) and Eq.~(\ref{etc28}), he can find the relation
between an interval of proper time $d\tau(R+x)$ at altitude $R+x$ and
an interval of proper time $d\tau(R)$ at altitude $R$
\begin{equation}
\label{etc29}
d\tau^2(R+x)=\left( 1-\frac{2\,\gamma\,R^2}{c^2(R+x)} \right)
\left( 1-\frac{2\,\gamma\,R^2}{c^2R} \right)^{-1} d\tau^2(R).
\end{equation}
Since $2\,\gamma\,R/c^2 \ll 1$ and $x \ll R$, the first order expansion
of formula~(\ref{etc29}) gives
\begin{equation}
\label{etc30}
d\tau^2(R+x) \approx \left( 1+\frac{2\,\gamma\,x}{c^2} \right)
d\tau^2(R).
\end{equation}

Let us now look at the case of an observer with a constant proper
acceleration. The formula~(\ref{etc5bis}) gives the interval of proper
time $d\tau(x)$ at altitude $x$ as a function of
an interval of proper time $dt$ at altitude $x=0$. For small values of
$x$ or for weak acceleration $A$, this formula can be written
\begin{equation}
\label{etc31}
d\tau^2(x) \approx \left( 1+\frac{2\,A\,x}{c^2} \right) d\tau^2(0).
\end{equation}
This relation is formally identical to relation~(\ref{etc30}). Locally,
the two metrics, the one for the gravitational field and the one for the
acceleration field, cannot be discriminated.

\section*{Acknowledgments}

The author (FNRS Research Associate) would like to thank the FNRS for
financial support.

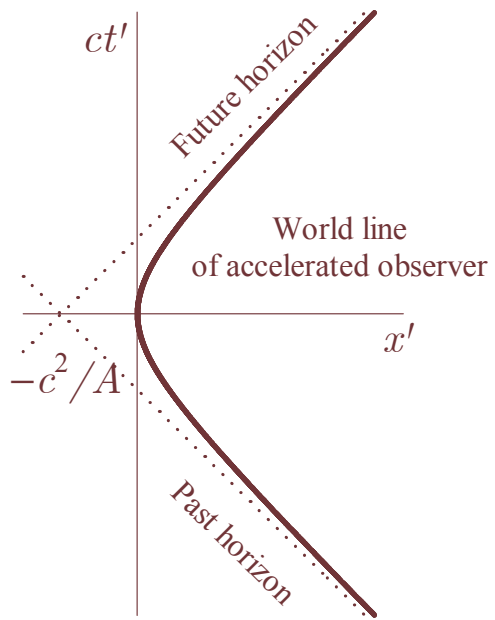
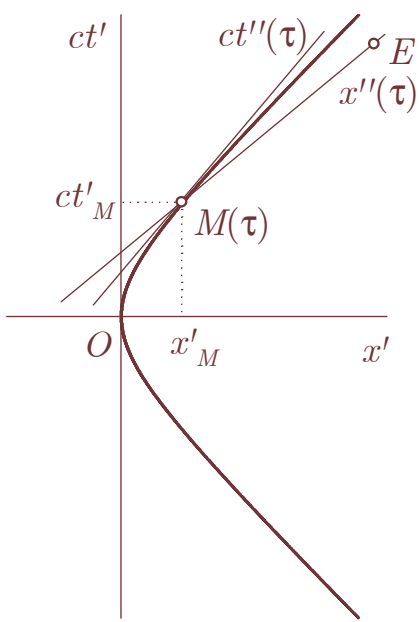
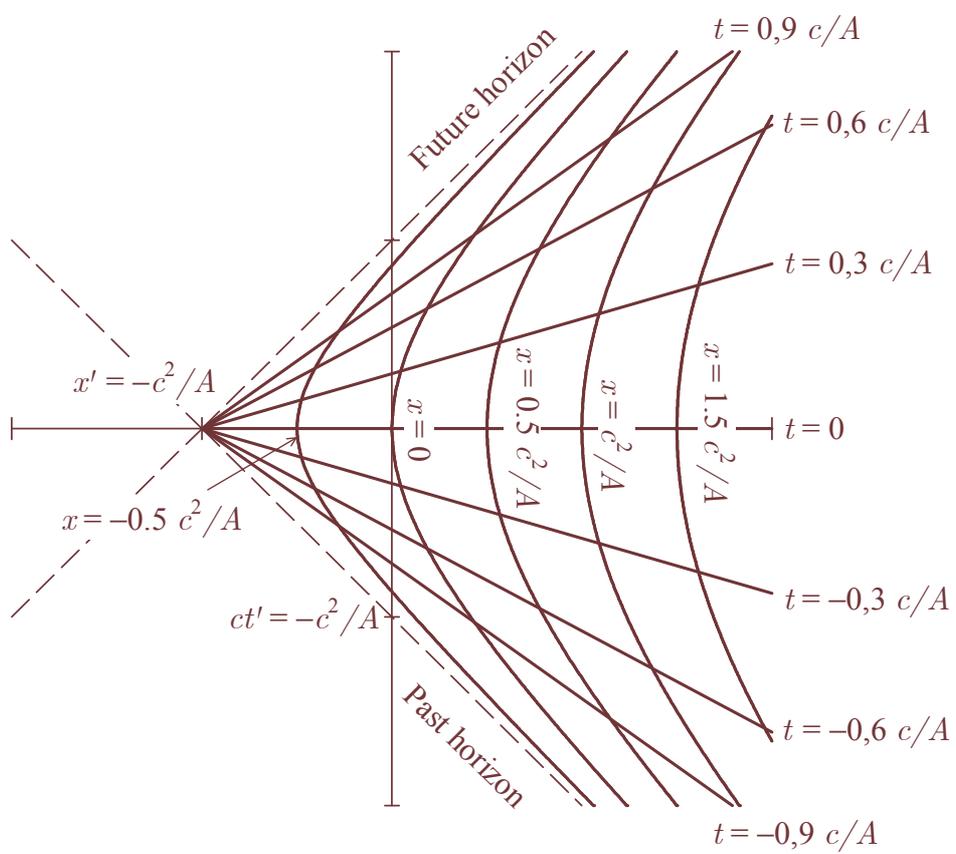
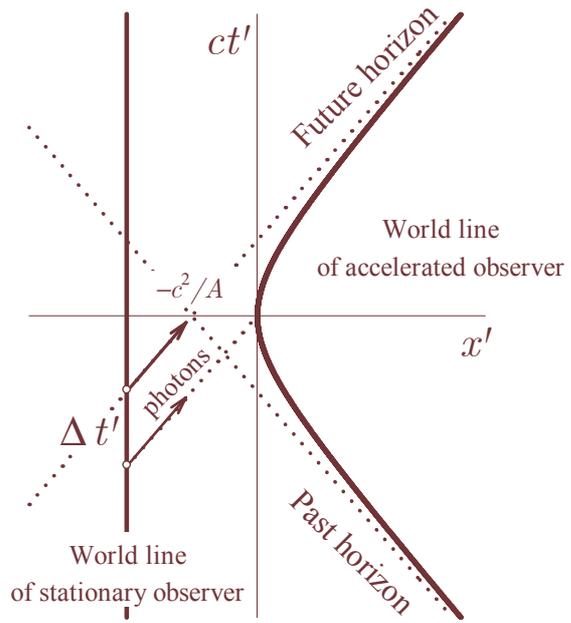
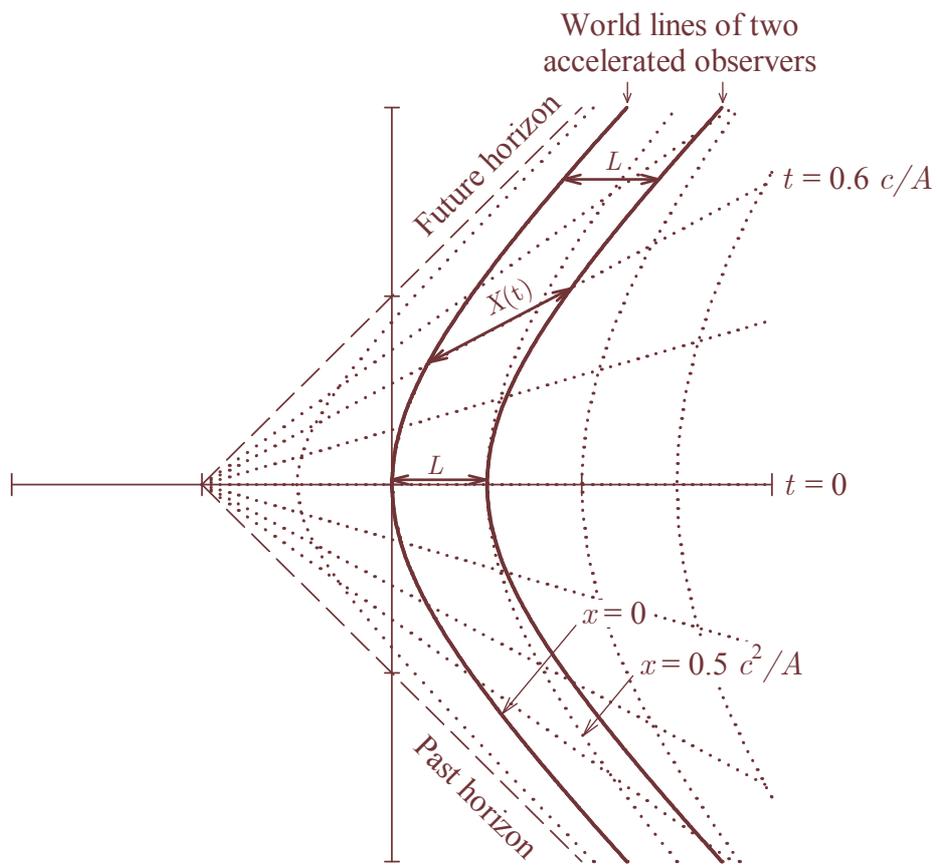
\end{document}